\documentclass{emulateapj} \usepackage{apjfonts}
\newcommand{\etal}{et~al.}
\newcommand{\MgIIdblt}{{\rm Mg}\kern 0.1em{\sc ii}~$\lambda\lambda 2796, 2803$}
\newcommand{\HI}{\hbox{{\rm H}\kern 0.1em{\sc i}}}
\newcommand{\Lya}{\hbox{{\rm Ly}\kern 0.1em$\alpha$}}
\newcommand{\Lyb}{\hbox{{\rm Ly}\kern 0.1em$\beta$}}
\newcommand{\MgII}{\hbox{{\rm Mg}\kern 0.1em{\sc ii}}}

\newcommand{\cmsq}{\hbox{cm$^{-2}$}}

\slugcomment{ApJ Lett.: Accepted Oct. 12 2012} 
\shorttitle{\sc {\MgII} Azimuthal Dependence}
\shortauthors{\sc Kacprzak et~al.}

\begin{document}


\title{Tracing Outflows and Accretion: A Bimodal Azimuthal Dependence of {\MgII} Absorption}


\author{\sc
Glenn G. Kacprzak\altaffilmark{1,2}
Christopher W. Churchill\altaffilmark{3}
and
Nikole M. Nielsen\altaffilmark{3}
}
                                                                                
\altaffiltext{1}{Swinburne University of Technology, Victoria 3122,
Australia {\tt gkacprzak@astro.swin.edu.au}}
\altaffiltext{2}{Australian Research Council Super Science Fellow}
\altaffiltext{3}{New Mexico State University, Las Cruces, NM 88003}

%%%% GLENN: I did not modify the abstract, but I think that it may require 
%%%% some adjustment by you if you adopt my Discussion on the EW distribution
%%%% Also, my estimate is 60% of all gas is wind gas (not 70% compared to 
%%%% accretion) as you have written

%%%% Please fix the statement ``The equivalent width distributions for gas 
%%%% along the major axis is smaller than for gas along the projected minor 
%%%% axis.''  - saying one distribution is smaller than another makes no sense.
%%%% you will need to say that the distribution is characterized by X being 
%%%% different in Y manner or say X is skewed toward higher/lower values 
%%%% relative to the other distribution, etc

\begin{abstract}

We report a bimodality in the azimuthal angle distribution of gas
around galaxies as traced by {\MgII} absorption: Halo gas prefers to
exist near the projected galaxy major and minor axes.  The bimodality
is demonstrated by computing the mean azimuthal angle probability
distribution function using 88 spectroscopically confirmed {\MgII}
absorption-selected galaxies [$W_r(2796)\geq0.1$~{\AA}] and 35
spectroscopically confirmed non-absorbing galaxies
[$W_r(2796)<0.1$~{\AA}] imaged with {\it HST\/} and SDSS.  The
azimuthal angle distribution for non-absorbers is flat, indicating no
azimuthal preference for gas characterized by $W_r(2796)<0.1$~{\AA}.
We find that blue star-forming galaxies clearly drive the bimodality
while red passive galaxies may exhibit an excess along their major
axis. These results are consistent with galaxy evolution scenarios
where star-forming galaxies accrete new gas, forming new stars and
producing winds, while red galaxies exist passively due to reduced gas
reservoirs.  We further compute an azimuthal angle dependent {\MgII}
absorption covering fraction, which is enhanced by as much as
20$-$30\% along the major and minor axes.  The $W_r(2796)$
distribution for gas along the major axis is likely skewed toward
weaker {\MgII} absorption than for gas along the projected minor
axis. These combined results are highly suggestive that the bimodality
is driven by gas accreted along the galaxy major axis and outflowing
along the galaxy minor axis. Adopting these assumptions, we find that
the opening angle of outflows and inflows to be $100^{\circ}$ and
$40^{\circ}$, respectively. We find the probability of detecting
outflows is $\sim$60\%, implying that winds are more commonly
observed.

\end{abstract}




\section{Introduction}
\label{sec:intro}
 
It is well established that {\MgIIdblt} absorption, detected in
background quasar/galaxy spectra, arises from gas associated with
foreground galaxies and thus provides a unique means to directly
observe mechanisms by which galaxies acquire, chemically enrich,
expel, and recycle their gaseous component \citep[see][for a
review]{cwc-china}.  The {\MgII} ion is ideal since it traces
metal-enriched low-ionization gas with $10^{16}\leq N(\HI)\leq
10^{22}$~{\cmsq} \citep{archiveI,weakII} and, as a result, is detected
out to projected galactic radii of $\sim 100$ kpc
\citep{kacprzak08,chen10a}. A significant quantity of {\HI} gas is
probed by {\MgII} absorption in galaxy halos with roughly 15\% of the
gas residing in DLAs and 5\% of the total hydrogen in stars
\citep{kacprzak11c,menard12}.

A large body of evidence suggests that {\MgII} absorption traces both
outflows from star-forming galaxies
\citep{bouche06,tremonti07,zibetti07,martin09,weiner09,chelouche10,nestor11,noterdaeme10,bordoloi11,coil11,rubin10,menard12,martin12}
and accretion onto host galaxies
\citep{steidel02,chen10a,chen10b,kacprzak10a,kacprzak11b,stewart11b,ribaudo11,kacprzak12,rubin12,martin12}.
While it is clear both processes are occurring, it is difficult to
disentangle which absorption systems may be uniquely associated with
either process. Since outflows are presumably metal-enriched, whereas
accreting material should be metal-poor, comparing the absorption-line
and host galaxy metallicity provides one suitable test for
discriminating winds from accretion \citep[e.g.,][and references
therein]{ribaudo11,kacprzak12}.  However it is not yet feasible to
perform this experiment for a large sample of galaxies.

The {\MgII} spatial distribution relative to their host galaxies also
provides a promising test.  Outflows are expected to extend along the
galaxy minor axis \citep[e.g.,][]{strickland04}, whereas accretion
would progress along filaments more planer to the galaxy
\citep[e.g.,][]{stewart11b}.  Modeling the morphologies of 40 {\it
HST} imaged absorbers, \citet{kacprzak11b} found that the rest-frame
{\MgII}~$\lambda 2796$ equivalent width, $W_r(2796)$, is dependent on
galaxy inclination, suggesting a co-planer geometry.  Furthermore, the
{\MgII} kinematics are consistent with being coupled to the galaxy
angular momentum \citep{steidel02,kacprzak10a}.  These results provide
evidence that some {\MgII} absorbing gas is accreting co-planer to
galaxies.

Stacking $\sim4000$ background galaxy spectra, \citet{bordoloi11}
showed that, statistically, unresolved {\MgII}-doublet equivalent
widths are larger along the galaxy projected minor axis than along the
projected major axis within $\sim 40$--$50$~kpc, suggesting that the
strongest absorption is ejected within winds along the minor axis.
Using 10 absorbers, \citet{bouche11} showed that {\MgII} absorption
arises along the projected minor and major axes of the host galaxies,
suggesting that {\MgII} is primarily detected in outflows and
accretion, however, the inclusion of a control sample of non-absorbers
is necessary to validate the bimodal claim.  Both works were further
validated by \citet{churchill12} who, using 65 galaxies, found that
{\MgII} absorption arises preferentially along the galaxy major axis
and then along the galaxy minor axis as $W_r(2796)$ increases; they
also found that for $W_r(2796)<0.1$~{\AA}, the galaxy inclinations and
azimuthal angles with respect to the quasar sight-line are consistent
with random distributions.

In view of these results, we aim to further explore the azimuthal
distribution of {\MgII} absorption for a large sample of
spectroscopically confirmed galaxies.  In this {\it Letter}, we
compute the mean azimuthal angle probability distribution function for
galaxies hosting $W_r(2796)\geq0.1$~{\AA} absorption and show that it
is bimodal, peaking near the galaxy minor and major axes.  We also
show that the distribution is flat for galaxies with
$W_r(2796)<0.1$~{\AA} (i.e., non-absorbers).  We further study this
distribution in terms of galaxy colors and equivalent widths.  We
adopt a $h=0.70$, $\Omega_{\rm M}=0.3$, $\Omega_{\Lambda}=0.7$
cosmology.

%Similarly, in stacked spectra of thousands of local star-
%forming galaxies from SDSS/DR7, Chen et al. (2010c) showed that the
%blue-shifted NaD I absorption is stronger within 60deg of the mi- nor
%axis.

\section{The Sample and Analysis}

We compiled a sample of 88 spectroscopically confirmed {\MgII}
absorption-selected galaxies with $W_r(2796)\geq0.1$~{\AA} [33 from
\citet{kacprzak11b} imaged with {\it HST}, 46 from \citet{chen10a} and
9 from \citet{kacprzak11a} imaged with SDSS] and 35 spectroscopically
confirmed non-absorbers with $W_r(2796)<0.1$~{\AA} [21 from
\citet{churchill12} imaged with {\it HST}, 14 from \citet{chen10a}
imaged with SDSS] having $0.09\leq z\leq1.1$ ($\left<z\right>=0.33$)
and $9\leq D\leq200$~kpc ($\left<D\right>=48$~kpc).  The galaxies
were selected to be isolated; group and double galaxies are not
included.  The $W_r(2796)=0.1$~{\AA} cut maximizes the number of
absorbers, while retaining a significant number of non-absorbers.

Galaxy AB magnitudes and colors were determined following the methods
of \citet{nielsen12}.  Rest-frame $B-K$ or $B-R$ were determined from
the measured apparent magnitudes. When only one color was measured, we
applied a conversion determined from a linear least-squares fit to the
average colors from galaxy spectral energy distribution templates of
\citet{mannucci01}. We obtained rest-frame colors for all but one
galaxy.

For this work, all galaxy morphological properties were obtained using
GIM2D \citep{simard02}.  56 galaxies were imaged with {\it HST}/WFPC2,
for which we adopt the morphologies and orientations measured by
\citet{kacprzak11b} and \citet{churchill12}.  For the
\citet{kacprzak11a} galaxies imaged with SDSS, we adopt the GIM2D
morphologies and orientations measured by \citet{kacprzak11a} and
\citet{bouche11}.

The SDSS galaxies taken from \citet{chen10a} have a median redshift of
$\left<z\right>=0.25$.  Given the short SDSS image exposure times
and the high redshifts, we studied the reliability of the GIM2D
models.  We identified 13 SDSS galaxies that had also been imaged with
{\it HST} from our previous work ($\left<z\right>=0.50$).  All SDSS
galaxies in our sample are resolved with $>$20 pixels at $1.5~\sigma$
above the background in the $r$-band.  Using the SDSS $r$-band images,
we computed azimuthal angles using GIM2D and then also from the
$1.5~\sigma$ isophotes using Source Extractor \citep{bertin96}. In
order to minimize the number of GIM2D free parameters, we fitted the
data using an exponential disk only, with zero bulge contribution.

We found that both the GIM2D and isophotal azimuthal angles of the
SDSS galaxies are consistent with the GIM2D azimuthal angles of the
{\it HST} galaxies (within $1~\sigma$ uncertainties of the {\it HST}
values).  In general, the errors on the GIM2D SDSS values are a factor
of a few larger, whereas it is difficult to quantify uncertainties for
the isophotal values.  Thus, we modeled the SDSS $r$-band imaged
galaxies as an exponential disk with GIM2D, applying the modeling
prescription of \citet{kacprzak11a}.  

We modeled all galaxy morphological types since 86\% of early-types
have regular stellar rotation (14\% exhibit slow rotation), spanning
the full range of apparent ellipticities, with 90\% being
kinematically and photometrically aligned
\citep[see][]{emsellem2011,krajnovic11}.

We adopt the convention of the azimuthal angle $\Phi=0^{\circ}$ to be
along the galaxy major axis and $\Phi=90^{\circ}$ to be along the
galaxy minor axis.

Direct binning of the azimuthal angles, the approach taken by
\citet{bouche11}, applies only when the measured uncertainties are
smaller than the bin size. Further complications arise when the
uncertainties are asymmetric, as is the case for GIM2D model
parameters. Here we take a different approach: we represent the
measured azimuthal angles and their uncertainties as univariate
asymmetric Gaussians \citep[see][]{kato02}, thus creating an azimuthal
angle probability distribution function (PDF) for each
galaxy\footnote{PDFs using the quadratic model of \citet{barlow03}
produce similar results, however their PDFs contain large unrealistic
spikes when the uncertainties are small.}.  From the continuous
azimuthal PDFs, we then compute the mean PDF as a function of $\Phi$.
The mean PDF represents the probability of detecting {\MgII}
absorption at a given $\Phi$.  This technique provides higher weight
per azimuthal angle bin for galaxies with well determined
$\Phi$. However, even the less robustly modeled galaxies provide
useful information; the method is equivalent to stacking low
signal-to-noise spectra or images to search for a coherent signal.

%%%%%%%%%%%%%%%%%%%%%%%%%%%%%%%%%%%%%%%%%%%%%%%%%%%%%%%%%%%%%%%%%%
\begin{figure}
\vglue -0.25in
\includegraphics[angle=0,scale=0.73]{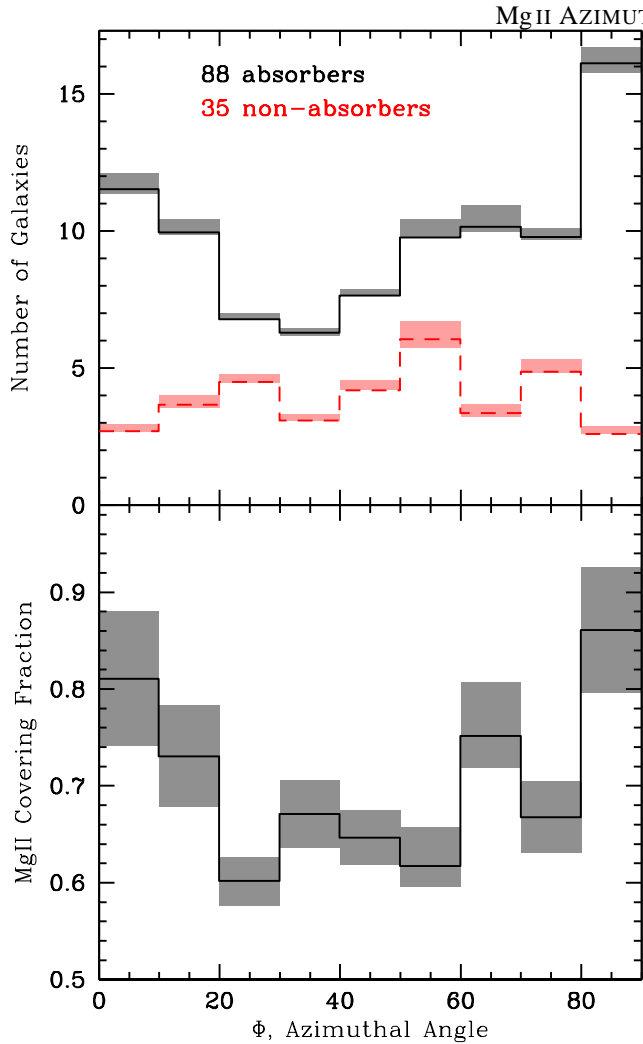}
\caption[angle=0]{(top) Binned azimuthal angle mean probability
distribution function for 88 absorbing galaxies (solid line) and 35
non-absorbing galaxies (dashed line). The areas of the histograms are
normalized to the total number of galaxies in each sub-sample. The
shaded regions are $1\sigma$ confidence intervals based upon a
jackknife analysis in which 10\% of the sub-sample was removed at
random. Given galaxies exist in multiple bins, the size of the
$1\sigma$ confidence intervals does not simply scale as the
square-root of the number of galaxies per bin. However, the sizes of
the confidence intervals will increase if more galaxies are removed in
the jackknife analysis.  The non-absorbers are consistent with a
random distribution within the expected Poisson noise.  The bimodal
distribution suggests a preference for {\MgII} absorbing gas toward
the galaxy major axis ($\Phi=0^{\circ}$) and along the minor axis
($\Phi=90^{\circ}$). --- (bottom) Azimuthal dependence of the covering
fraction with shaded regions providing the $1\sigma$ confidence
intervals. Note the higher covering fractions along the major and
minor axes.}
\label{fig:main}
\vglue -0.05in
\end{figure}
%%%%%%%%%%%%%%%%%%%%%%%%%%%%%%%%%%%%%%%%%%%%%%%%%%%%%%%%%%%%%%%%%%

\section{The Bimodal Azimuthal Distribution}

In Figure~\ref{fig:main}, we present the binned mean azimuthal
angle PDF for the 88 absorbing and 35 non-absorbing galaxies. The
binned PDFs have been normalized to the total number of galaxies in
each sub-sample.  The shaded regions about each bin are the $1~\sigma$
deviations produced by a jackknife analysis in which we randomly
removed 10\% of the sample for a million permutations.

The non-absorbers exhibit a relatively flat distribution, as expected
for random $\Phi$, and the fluctuations are consistent with Poisson
noise. This result confirms the work of \citet{churchill12}, who
showed that the azimuthal angle distribution is random for
$W_r(2796)<0.1$~{\AA}.

The absorbing galaxies exhibit a bimodal $\Phi$ distribution, with one
peak corresponding to the background quasars probing the galaxy
projected major axis and a second, higher and broader peak, when the
galaxy is probed along the projected minor axis.  The shape of the
mean PDF indicates that {\MgII} absorbing gas is preferentially
located along the projected major and minor axes of galaxies.  Our
result confirms the innovatory work of \citet{bouche11} based upon 10
galaxies.  A plausible scenario is that wind/outflow gas is
distributed about the galaxy minor axis with relatively high
frequency, whereas accreting/infalling gas is frequently found in a
co-planer geometry.  For the remainder of the discussion, we refer to
major axis absorption as infalling or accreting gas and to the minor
axis absorption as outflowing or wind gas.

We further divided our sample of absorbing galaxies into two impact
parameter sub-samples, those with $D>40$ kpc and $D<40$ kpc, and found
that the bimodal shape of the mean PDF is well preserved for both
impact parameter sub-samples.  This implies that outflowing and
accreting gas maintain their major and minor axis orientation
preferences, respectively, far out into the halo.

The width of each peak in the mean PDF provides rough constraints on
the geometry of outflowing and inflowing {\MgII} gas.  The peak at
$\Phi=0^{\circ}$ suggests that accreting gas is found within
$\Delta\Phi\simeq\pm20^{\circ}$ of the galaxy plane.  For
$\Phi=90^{\circ}$, the peak is initially narrow, suggesting the
majority of galaxies expel gas within $\Delta\Phi\simeq\pm10^{\circ}$
of their minor axis.  The much broader tail of the $\Phi=90^{\circ}$
peak suggests that the opening angle of outflowing gas may be as large
as $\Delta\Phi\simeq\pm50^{\circ}$, but also the frequency at which
this gas is detected decreases for larger opening angles.  Our
estimated outflow opening angle is consistent with other published
values determined using different techniques
\citep[e.g.,][]{walter02,martin12}.  The ratio of the area between the
outflowing and infalling gas, bifurcated at $40^{\circ}$, suggests
that $\simeq60$\% of {\MgII} absorbing gas is outflowing.

In Figure~\ref{fig:main}, we plot the {\MgII} gas covering fraction,
which is the ratio of the number of absorbers to the sum of absorbers
and non-absorbers in each azimuthal bin.  This first presentation of
the covering fraction as a function of azimuthal angle shows a peak at
0.8--0.9 along the projected minor axis, a decrease of 20--30\% at
intermediate $\Phi$, and an increase again toward the projected major
axis.  The mean covering fraction of our sample, 72\%, is consistent
with previous results \citep[e.g.,][]{kacprzak08,chen10a}.

%%%%%%%%%%%%%%%%%%%%%%%%%%%%%%%%%%%%%%%%%%%%%%%%%%%%%%%%%%%%%%%%%%
\begin{figure}
\vglue -0.25in
\includegraphics[angle=0,scale=0.76]{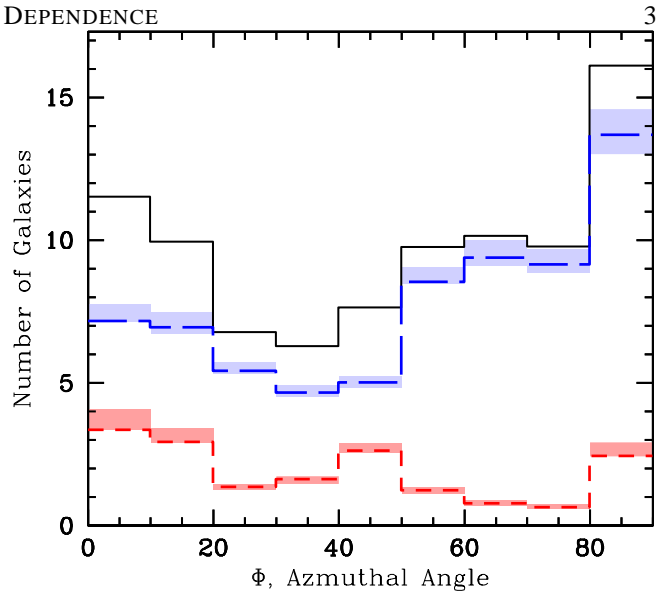}
\caption[angle=0]{Galaxy color dependence of the azimuthal
distribution of {\MgII} absorbers. The solid line (black) is the
distribution shown in Figure~\ref{fig:main}. The color selection
adopted from \citet{chen10a} of $B-R\leq1.1$ represents late-type
galaxies (dashed blue line) and $B-R>1.1$ represents early-type
galaxies (dotted red line). Shaded regions are $1\sigma$ confidence
intervals based upon a jackknife analysis. The data are consistent
with star-forming galaxies being dominated by outflows. }
\label{fig:colors}
\vglue -0.05in
\end{figure}
%%%%%%%%%%%%%%%%%%%%%%%%%%%%%%%%%%%%%%%%%%%%%%%%%%%%%%%%%%%%%%%%%%

%%%%%%%%%%%%%%%%%%%%%%%%%%%%%%%%%%%%%%%%%%%%%%%%%%%%%%%%%%%%%%%%%%
\begin{figure*}
\centering
\begin{tabular}{cc}
\includegraphics[angle=0,scale=0.58]{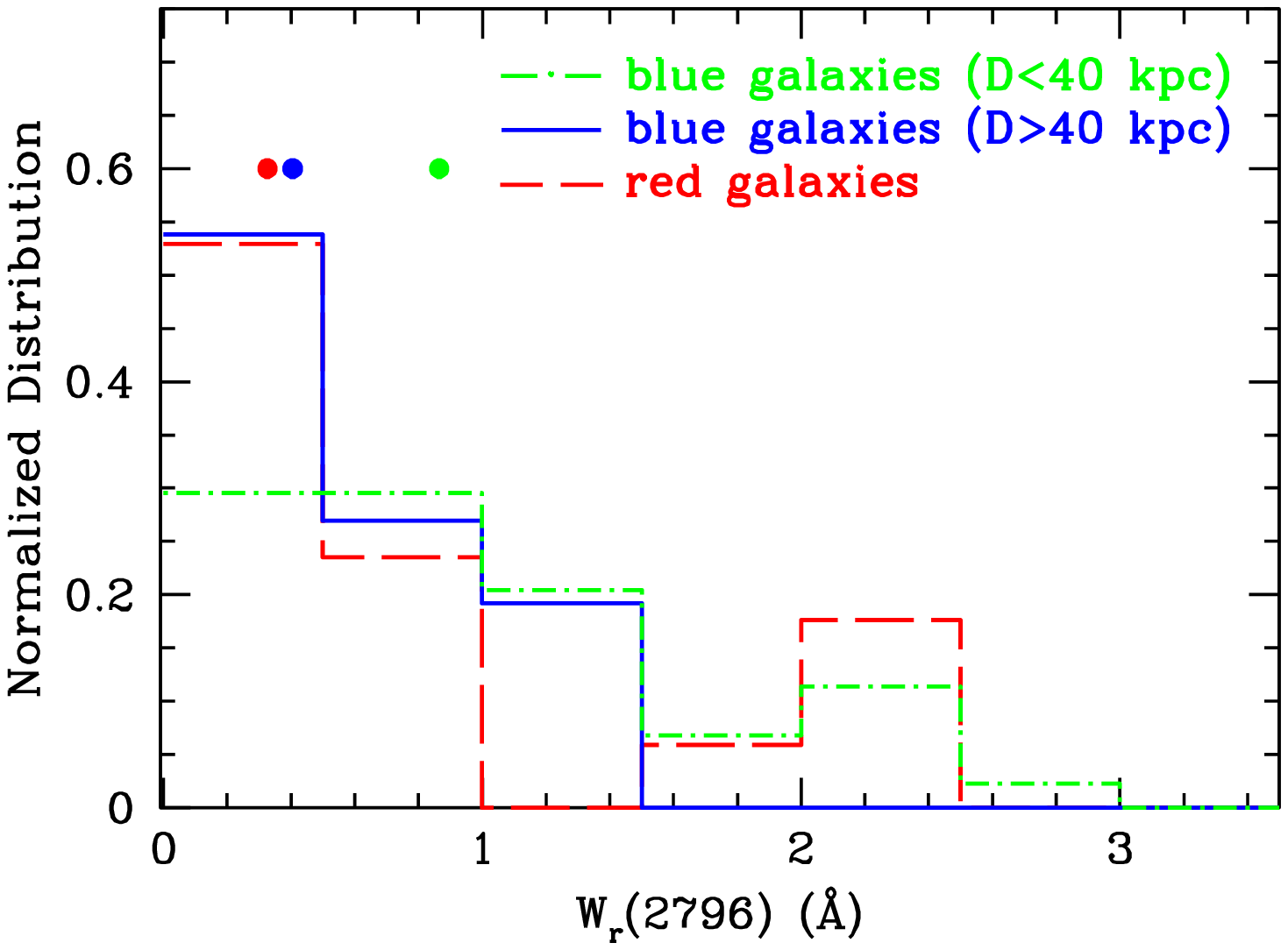} &
\includegraphics[angle=0,scale=0.58]{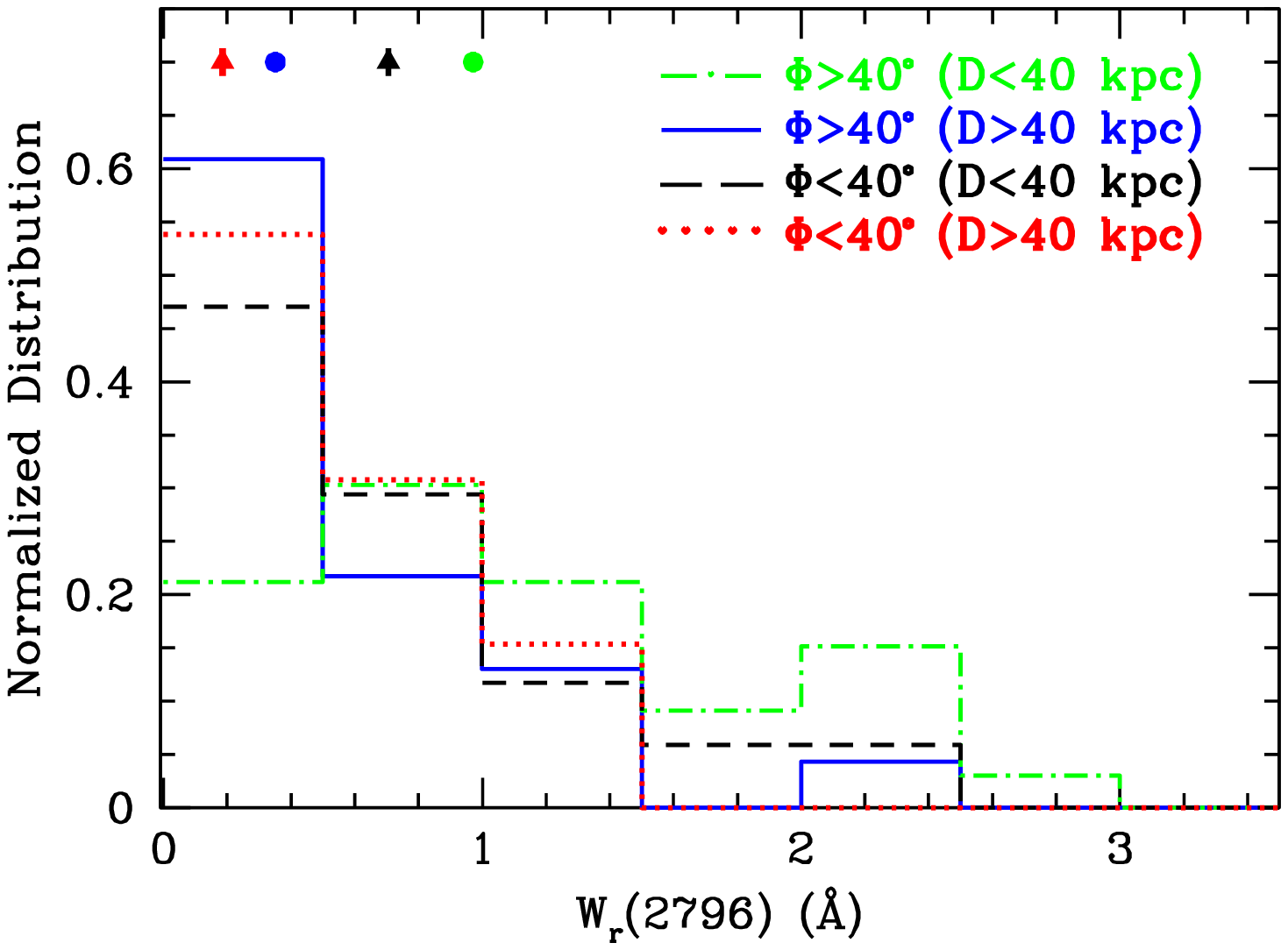} \\
\end{tabular}
  \caption{ (left) The area-normalized distribution of $W_r(2796)$ for
  blue galaxies with $D<40$~kpc, blue galaxies with $D>40$~kpc, and
  red galaxies. The weighted mean of each distribution is shown with
  the solid points above the histogram; the errors are equivalent to
  the point size. While consistent with \citet{bordoloi11}, we found
  blue galaxies at lower $D$ having larger $W_r(2796)$ compared to red
  galaxies and blue galaxies at $D>40$~kpc. --- (right) The
  $W_r(2796)$ dependence for $\Phi>40^{\circ}$ and $\Phi<40^{\circ}$
  separated into two $D$ bins. The $W_r(2796)$ weighted mean for each
  sub-sample is shown above the distributions with $\Phi>40^{\circ}$
  indicated by circles and $\Phi<40^{\circ}$ by arrows.  The data
  suggest a difference in the $W_r(2796)$ distributions as a function
  of both $D$ and $\Phi$.  This may imply that either there is less
  gas column density along the projected galaxy major axis then along
  the projected minor axis, {\it or}, the gas could be metal-enriched
  along the galaxy minor axis compared to the major axis. This would
  be expected in a accretion/wind scenario. }
  \label{fig:ew}
\end{figure*}
%%%%%%%%%%%%%%%%%%%%%%%%%%%%%%%%%%%%%%%%%%%%%%%%%%%%%%%%%%%%%%%%%%

\subsection{Colors, Equivalent Widths, and $\Phi$}

We separated the sample of absorbers into early-type and late-type
galaxies based on their rest-frame $B-R$ colors.  Following
\citet{chen10a}, we adopted the color cut $B-R\leq1.1$ to represent
late-type (blue) galaxies and $B-R>1.1$ to represent early-type (red)
galaxies.  In Figure~\ref{fig:colors}, we present the mean azimuthal
angle PDF broken down by galaxy color. The early-type galaxies exhibit
a relatively flat distribution with $\sim\pm1~\sigma$ fluctuations
across bins. However, the elevated frequency at $\Phi=0^{\circ}$ is
suggestive that gas is accreted along the major axis, which may
provide the cool gas reservoirs observed around early-type galaxies
\citep[e.g.,][]{grossi09}.  The late-type galaxies clearly dominate
the overall bimodal distribution.  This result is intuitive if we
speculate that red galaxies have significantly less accreting gas, and
therefore little fuel for star formation, resulting in no outflowing
gas.  For blue galaxies, the accretion is higher, providing fuel for
star formation that then produces outflows.  We also split the
absorber sample by galaxy color to yield equal numbers of red and blue
galaxies in each sub-sample; the same trend appears.  The azimuthal
bimodality of blue galaxies and the flat distribution of red galaxies
is consistent with the picture in which blue star-forming galaxies
exhibit outflows due to accreted gas fueling star formation.

One might argue that red galaxies may have a preference for weaker
absorption \citep[see][]{zibetti07,bordoloi11}.  We thus investigated
whether the $W_r(2796)$ distribution depends on galaxy color and
$D$. In Figure~\ref{fig:ew}, we show the area-normalized distribution
of $W_r(2796)$ for red galaxies and for blue galaxies with $D<40$ kpc
and $D>40$ kpc (we have too few red galaxies to separate them into
impact parameter bins).  The blue galaxies at smaller $D$ have a
higher optimally weighted mean $W_r(2796)$ than blue galaxies at
larger $D$.  This is consistent with the well-known anti-correlation
between {\MgII} equivalent width and impact parameter \citep[cf.,][and
references therein]{nielsen12}.  The sub-sample of red galaxies has a
mean $W_r(2796)$ consistent with the blue galaxies at larger $D$.

Interestingly, Figure~\ref{fig:ew} shows a paucity of
$W_r(2796)\leq0.5$~{\AA} absorption at smaller $D$ for blue galaxies
compared to both blue galaxies at larger $D$ and to all red galaxies.
Also, note the paucity of $W_r(2796)\geq1.5$~{\AA} absorption at
larger $D$ for blue galaxies as compared to both blue galaxies at
smaller $D$ and to all red galaxies.  However, a KS test yields that
the $W_r(2796)$ distribution of blue galaxies at smaller $D$ are
statistically consistent with the $W_r(2796)$ distributions of both
the red galaxies and the blue galaxies at larger $D$ (2.0~$\sigma$, a
high significance level is difficult to achieve with small number
statistics).

The suggested weaker absorption associated with red galaxies and with
blue galaxies at larger $D$ in our sample is consistent with the
findings of \citet{bordoloi11}, who interpret this as winds extending
out to projected distances of $\sim 40$--$50$~kpc.  However, the
bimodality in the azimuthal distribution is present for our $D>40$~kpc
absorber sample, suggesting winds persist beyond $40$~kpc, but with
smaller $W_r(2796)$ indicating that the wind gas thins out at larger
$D$.

We further investigated whether the distribution of $W_r(2796)$
differs for infalling gas or outflowing gas by splitting the
$D<40$~kpc and $D>40$~kpc absorbers into the two azimuthal bins,
$\Phi<40^{\circ}$ for inflowing gas and $\Phi>40^{\circ}$ for
outflowing gas.  The value $\Phi=40^{\circ}$ is the point of
inflection in the mean azimuthal angle PDF.  In Figure~\ref{fig:ew},
we present the area-normalized $W_r(2796)$ distributions.  For
$\Phi>40^{\circ}$, the mean $W_r(2796)$ differ by a factor of 3.3 for
the smaller and larger $D$ sub-samples.  Similarly, for
$\Phi<40^{\circ}$, the mean $W_r(2796)$ differ by a factor of 3.8 for
the smaller and larger $D$ sub-samples.  The $W_r(2796)$ distribution
for outflowing gas at smaller $D$ is characterized by a paucity of
$W_r(2796)\leq0.5$~{\AA} absorption and a higher abundance of
$W_r(2796)\geq1.5$~{\AA} absorption, whereas the $W_r(2796)$
distribution for outflowing gas at larger $D$ is characterized by an
abundance of $W_r(2796)\leq0.5$~{\AA} absorption and a paucity of
$W_r(2796)\geq1.5$~{\AA} absorption. A KS test yields that the
populations differ at the $2.6\sigma$ significance level.

Comparing the $D<40$ kpc $W_r(2796)$ distributions for wind versus
accreting gas, we note a similar trend; at smaller impact parameters,
accreting gas exhibits a higher abundance of $W_r(2796)\leq0.5$~{\AA}
absorption and a paucity of $W_r(2796)\geq1.5$~{\AA} absorption. KS
tests yield that the outflowing gas with $\Phi>40^{\circ}$ and
$D<40$~kpc differs from the inflowing gas with $\Phi<40^{\circ}$ and
$D>40$~kpc at the $3.1\sigma$ significance level.  Qualitatively,
these combined results may imply that either there is less column
density and/or velocity spread of gas in projection along the galaxy
major axis than along the minor axis, {\it and/or}, the gas may be
more metal-enriched along the galaxy minor axis than along the galaxy
major axis.  These trends would be expected for the accretion/wind
scenario we have described.

%outflowing gas with $\Phi >
%40^{\circ}$ and $D<40$~kpc differs from the outflowing gas with $\Phi
%> 40^{\circ}$ and $D>40$~kpc at the $2.6\sigma$ significance level,
%while 

\section{Conclusion}

We have demonstrated that the distribution of {\MgII} absorption
around galaxies exhibits an azimuthal angle bimodality when compared
to non-absorbing galaxies, whereby {\MgII} absorption is preferred
near the projected major and minor axes. Outflows and accretion likely
extend beyond 40~kpc, but with smaller mean $W_r(2796)$.  Furthermore,
the blue star-forming galaxies drive the bimodality, suggesting that
the accretion of gas drives star-formation that produce outflows.  Red
galaxies may exhibit some gas accretion along the major axis. Under
these assumptions, we compute opening angles of outflows and inflows
to be $100^{\circ}$ and $40^{\circ}$, respectively.

In addition, the data suggest that star-forming galaxies at low impact
parameters exhibit higher mean $W_r(2796)$ than red galaxies and
star-forming galaxies with larger impact parameters.  Also, the
$W_r(2796)$ distribution for the infalling gas along the major axis
exhibits a smaller mean $W_r(2796)$ than outflowing gas along the
minor axis.  This could result from less gas being probed along the
major axis than the minor axis, less velocity structures/dispersions
expected for winds, or higher metal enrichment along the minor axis,
as expected for wind driven material.  The probability of detecting
outflows is $\sim$60\%, implying that winds are more commonly
observed, likely because the opening angle of outflows is 2.5 times
larger than for accreting gas.

The data paint a picture that is consistent with the general idea of
how the buildup and evolution of galaxies occurs as well as how
galaxies enrich their gaseous halos.  Our results signify that
additional {\it HST\/} imaging of absorbing and non-absorbing galaxies
will substantially increase our understanding of galactic scale
feedback and accretion of intergalactic gas.

%%%%%%%%%%%%%%%%%%%%%%%%%%%%%%%%%%%%%%%%
\acknowledgments 

We thank David Law for additional data.
% CWC and NMN were supported through grant HST-GO-11667.01-A provided by
% NASA via the Space Telescope Science Institute, which is operated by
% the Association of Universities for Research in Astronomy (AURA) under
% NASA contract NAS 5-26555.
CWC and NMN were supported through NASA/STScI grant HST-GO-11667.01-A
and NASA's New Mexico Space Grant Consortium.  Based on Hubble Legacy
Archive data.  Based on data from W.M. Keck Observatory, a scientific
partnership between the Caltech, the University of California, and
NASA. %the National
% Aeronautics and Space Administration.
%%%%%% The web site says this is optional, and we need room
% The Observatory was made
%possible by the generous financial support of the W.M. Keck
%Foundation.  
%Data was also from The Sloan Digital Sky Survey (SDSS).  Funding for
%SDSS/SDSS-II has been provided by the Alfred P. Sloan Foundation, the
%Participating Institutions, the National Science Foundation, the
%U.S. Department of Energy, NASA, the Japanese Monbukagakusho, the Max
%Planck Society, and the Higher Education Funding Council for England.
Data from The Sloan Digital Sky Survey (SDSS/SDSS-II), which is funded
by the Alfred P. Sloan Foundation, Participating Institutions, NSF,
U.S. Department of Energy, NASA, Japanese Monbukagakusho, Max Planck
Society, and the Higher Education Funding Council for England.
%The SDSS Web Site is http://www.sdss.org/. The SDSS is
%managed by the Astrophysical Research Consortium for the Participating
%Institutions.

%% Included in this acknowledgments section are examples of the
%% AASTeX hypertext markup commands. Use \url without the optional [HREF]
%% argument when you want to print the url directly in the text. Otherwise,
%% use either \url or \anchor, with the HREF as the first argument and the
%% text to be printed in the second.

%doing the math in section~\ref{bozomath}.
%More information on the AASTeX macros package is available \\ at
%\url{http://www.aas.org/publications/aastex}.
%For technical support, please write to
%\email{aastex-help@aas.org}.


%% in parentheses, after the keyword, but they will not be verified.

{\it Facilities:} \facility{HST (WFPC--2)}, \facility{Keck I (HIRES,
LRIS)}, \facility{VLT (UVES)}, \facility{Sloan (SDSS)}.

%\newpage
%\begin{figure}
%\vglue -0.25in
%\includegraphics[angle=0,scale=0.35]{/nfs/cluster/qso/gkacprzak/SDSSinc/Errorcode/plots/PAvsD.eps}
%\caption[angle=0]{The $\Phi$ distribution for D>40kpc and D<40 kpc. The bimodality still survives}
%\label{fig:nW}
%\vglue -0.05in
%\end{figure}
%
%\begin{figure}
%  \epsscale{0.9}\plottwo{/nfs/cluster/qso/gkacprzak/SDSSinc/Errorcode/plots/PAandEW.eps}{/nfs/cluster/qso/gkacprzak/SDSSinc/Errorcode/plots/BKewsChen.eps}
%  \caption{Top EW dependence for $\Phi> 40$ and $\Phi<
%  40$. KS(P)=1.92E10-1, thus similar distribution. EW dependence on
%  galaxy color. KS(P)=5.629E-1, thus similar distribution.  }
%  \label{maps}
%\end{figure}
%
%\begin{figure}
%\vglue -0.25in
%\includegraphics[angle=0,scale=0.35]{/nfs/cluster/qso/gkacprzak/SDSSinc/Errorcode/plots/PAgt40EWvsi.eps}
%\caption[angle=0]{PDFs for galaxy inclination when $\Phi>40$ degrees
%and the dependence on EW. The distributions are the same, however,
%there is a preference for High inclined galaxies. This is a
%cross-section argument given that the largest cross-section for winds on
%the plan of the sky is for edge-on galaxies. The Wind cross-section
%decreases as one goes to i>90. The THICK line is the non-absorbers with $\Phi>40$. the distribution is
%flat, though dips for face-on galaxies due to low number statistics.}
%\label{fig:nW}
%\vglue -0.05in
%\end{figure}

\end{document}